\documentclass[12pt,aps,showpacs,nofootinbib,showkeys]{revtex4-2}
\usepackage{amsmath}    % need for subequations
\usepackage{graphicx}   % need for figures
\usepackage{color}      % use if color is used in text
\usepackage{amssymb}
\usepackage{setspace}
\usepackage{amsfonts}

\begin{document}
%\setstretch{1.25}

% Keywords command
%\providecommand{\keywords}[1]
%{
%  \small	
%  \textbf{\textit{Keywords---}} #1
%}

%\newcommand{\blue}[1]{\textcolor{blue}{#1}}
%\newcommand{\new}{\blue}
%\newcommand{\green}[1]{\textcolor{green}{#1}}
%\newcommand{\modif}{\green}
%\newcommand{\red}[1]{\textcolor{red}{#1}}
%\newcommand{\attention}{\red}

%Color scheme: {\color{green} green: comments about something along
%the text,}~~{\color{blue} blue: replacing the original text.}
%{\color{red} red: wrong idea or approach.}

\title{Gravitational Surface Tension as the Origin for the Black Hole Entropy}

\author{S. D.  Campos}\email{sergiodc@ufscar.br}

\author{R. H. Longaresi}

\address{Applied Mathematics Laboratory}

\address{GDTeC - Development Group for Educational Technologies in Sciences - CCTS/DFQM, Federal University of S\~ao Carlos, Sorocaba, CEP 18052780, Brazil}

\begin{abstract}
In this work, we explore the thermodynamics of black holes using the Gouy-Stodola theorem, traditionally applied to mechanical systems relating entropy production to the difference between reversible and irreversible work. We model black holes as gravitational bubbles with surface tension defined at the event horizon, deriving the Bekenstein-Hawking entropy relation for non-rotating black holes. One extends this approach to rotating black holes, incorporating the effects of angular momentum, demonstrating that the Gouy-Stodola theorem can similarly derive the entropy-area law in this case. Additionally, we analyze the merging of two black holes, showing that the resultant total entropy exceeds the sum of the individual entropies, thereby adhering to the second law of thermodynamics. Our results suggest that gravitational surface tension is a key factor in black hole thermodynamics, providing a novel and coherent framework for understanding the entropy production in these extreme astrophysical objects.
\end{abstract}




%\date{\today}


\maketitle
%%%%%%%%%%%%%%%%%%%%%%%%%%%%%
\section{Introduction}

The physics of black holes (BH) appears to be exotic to common sense, pushing physical concepts to their limits. These strange entities appear to follow their own laws of thermodynamics, but currently, the best we can provide is an analogy \cite{J.M.Bardeen.B.Carter.S.W.Hawking.Commun.Math.Phys.31.161.1973} (criticism can be seen in Ref. \cite{J.Dougherty.C.Callender}).

The laws of thermodynamics are settled down on mathematics/physics arguments and experimental evidence. However, when discussing black holes, experimental evidence about what occurs at their core is obstructed by the no-hair conjecture \cite{W.Israel.Phys.Rev.164.5.1776.1967,W.Israel.Commun.Math.Phys.8.3.245.1968,b.carter.phys.rev.lett.26.6.331.1971}: only mass, charge, and angular momentum can describe the exterior of the black hole, and all internal information cannot be recovered using these three classical observables.

In trying to improve the understanding of BH thermodynamics, one assumes that a BH can be naively modeled as a soap bubble.  Indeed, one assumes the black hole is a gravitational bubble centered in the singularity. In the simple case of a non-rotating and non-charged BH, the gravitational surface tension is defined by the surface at the Schwarzschild radius, the event horizon surface. It is important to stress the gravitational bubble presented here is not the compact form with zero Weyl tensor \cite{V.A.Berezin.V.I.Dokuchaev.Yu.N.Eroshenko.IJMPA.31.2.1641004.2016} or a foam whose basic unit is a gravitational bubble \cite{S.W.Hawking.D.N.Page.C.N.Pope}. The gravitational bubble here corresponds to the gravitational surface defined by the surface delimited by the event horizon radius. For a static and spherically symmetric black hole, this radius plays the role of a radial measure.

The surface stress tensor that defines the gravitational surface is an old central problem in gravitation \cite{W.Israel.Nuvo.Cimmento.B.44.1.1966,P.Beltracchi.P.Gondolo.E.Mottola.Phys.Rev.D.105.024001.2022}. For a simple bubble, the junction conditions about the two different geometries of the inside and outside spacetime are determinant for the study of first-order phase transitions. For a black hole, the event horizon is viewed as a coordinate singularity in the sense it can be removed by adopting a suitable change of coordinates. However, the no-hair theorem ensures the event horizon surface yields real thermodynamic physical consequences. 

There is an almost unknown result in physics called Gouy-Stodola theorem \cite{A.Stodola.Zeitschr.d.Verein.Deutscher.Ingenieure.32.1086.1898,G.Gouy.J.dePhys.8.35.1899}, that attributes the entropy produced in a mechanical system to the difference between the work due to reversible and irreversible processes \cite{longaresi.campos.1,longaresi.campos.2} occurring in such a system. Disregarding the charge and using the Gouy-Stodola theorem, one obtains the entropy produced for two systems: a non-rotating black hole and a rotating one. In the first case, one derives the usual Hawking-Bekenstein relation between entropy and the event horizon surface \cite{S.W.Hawking.Commun.Math.Phys.43.199.1975,J.D.Bekenstein.Phys.Rev.D7.2333.1973}. Further, for the rotating black hole, one obtains the same result as stated for a non-charged black hole \cite{J.M.Bardeen.B.Carter.S.W.Hawking.Commun.Math.Phys.31.161.1973}.

One discusses the merging of two black holes from the point of view of the Gouy-Stodola theorem under the gravitational bubble assumption. One obtains the total entropy greater than the sum of individual entropy of each BH.  

The present manuscript is structured as follows. In section \ref{sec:section_0}, one introduces briefly the Gouy-Stodola theorem. Section \ref{sec:section_1} presents the entropy-area law. In section \ref{sec:section_bubble}, one introduces the gravitational bubble model and presents our results for non-rotating and rotating black holes. In section \ref{sec:merging}, one discusses the entropy emergence in the merger of two black holes. Discussion is left to the final section \ref{sec:disc}.

%%%%%%%%%%%%%%%%%%
\section{The Gouy-Stodola Theorem Applied to Black Holes}\label{sec:section_0}

The Gouy-Stodola theorem \cite{G.Gouy.J.dePhys.8.35.1899,A.Stodola.Zeitschr.d.Verein.Deutscher.Ingenieure.32.1086.1898}, with its thermodynamic origin, is a powerful tool that allows for the calculation of entropy produced in mechanical systems. A pedagogical development can be seen in Ref. \cite{longaresi.campos.1}, while an application for entropy produced in the inflationary scenario of the Universe is presented in Ref. \cite{longaresi.campos.2}. 

This theorem states that, in an isotherm process,  the entropy yielded ($\dot{S}$) in a mechanical system can be obtained by \cite{longaresi.campos.1,longaresi.campos.2}
\begin{eqnarray}\label{eq:gouysotodola}
    T\dot{S}=\dot{W}_r-\dot{W}_i,
\end{eqnarray}
\noindent where $\dot{W}_r$ and $\dot{W}_i$ are, respectively, the time derivative of the work due to the conservative and non-conservative forces that act in the system. This theorem relates entropy production to the work done by the non-conservative force and thus provides a practical way to obtain entropy production in a physical system in both senses: qualitative and quantitative. The presence of irreversibility leads to a time decrease in energy implying $\delta \dot W_{i}<\delta \dot W_{r}$, consistent with the fact that $\dot S$ is always positive.

There are two observations one can present here. The first one is that a BH is not a mechanical system. Indeed, it belongs to the highly relativistic and quantum domain. However, the forces acting on this system can be classified into two cases: conservative and non-conservative. The work associated with each force can be calculated. The second observation is that theoretically speaking, we can access and recognize only forces acting outside the event horizon of the black hole. The forces acting inside the event horizon are completely unknown. However, the BH evaporation may indicate the presence of irreversible processes inside the BH, leading to the supposition that non-conservative forces are acting inside the BH.

In the present model, one circumvents the problem of knowing exactly the forces (inside or outside) by using the gravitational bubble model, as shall be seen further.

%%%%%%%%%%%%%%%%%%
\section{The Entropy-Area Law for Black Hole}\label{sec:section_1}

This section assumes a non-rotating, non-charged, spherically symmetric BH for simplicity. In this case, the entropy-area law states that the entropy of a BH is given by the Bekenstein-Hawking result \cite{J.D.Bekenstein.Phys.Rev.D7.2333.1973,S.W.Hawking.Commun.Math.Phys.43.199.1975}
\begin{eqnarray}\label{eq:bekhaw}
    S_{bh}=\frac{k c^3}{4G\hbar} A_{eh},
\end{eqnarray}
\noindent where $A_{eh}$ is the event horizon surface, $k$ is the Boltzmann constant, $c$ is the velocity of light, $G$ is the gravitational constant, and $\hbar$ is the Planck constant divided by $2\pi$. It should be emphasized that $S_{bh}$ does not necessarily correspond to the {\it thermodynamic entropy} of the BH. By analogy, it was assumed that the event horizon surface \textit{behaves} as an entropy because of its mathematical properties: the greater the BH (event horizon), the greater the entropy \cite{J.D.Bekenstein.Phys.Rev.D7.2333.1973,S.W.Hawking.Commun.Math.Phys.43.199.1975}. The same analogy was extended to the other laws governing BH thermodynamics \cite{J.M.Bardeen.B.Carter.S.W.Hawking.Commun.Math.Phys.31.161.1973}

Bearing in mind the analogy with thermodynamics, note that the factor $kc^3/4G\hbar$ can be written as 
\begin{eqnarray}
    \frac{kc^3}{4G\hbar}=\frac{\sigma}{T},
\end{eqnarray}
\noindent where $T$ is the temperature for an isothermic process and using the standard values for the fundamental constants, one has $\sigma= 1.32137\times10^{46}$ kg/s$^2$ (or N/m). Due to its units, $\sigma$ can be interpreted as the surface tension or surface free energy \cite{Adamson.book} associated with the surface of the BH given by $A_{eh}$.

In the general context of fluids, surface tension emerges as an interfacial temperature-dependent phenomenon, where the asymmetry of forces between the internal and external sides induces the surface to work as a stretched (elastic) membrane. In the present context, the surface tension $\sigma$ for the BH may indicate the existence of a region of the space with similar properties to a membrane but with gravitational origin. 

Considering the above comments, the Bekenstein-Hawking entropy law can be rewritten as
\begin{eqnarray}
    S_{bh}=\frac{k c^3}{4G\hbar}A_{eh}=\frac{1}{T}\sigma A_{eh},
\end{eqnarray}
\noindent where for a spherically symmetric BH, the event horizon surface is given by
\begin{eqnarray}\label{eq:eventarea}
    A_{eh}=4\pi r_s^2,
\end{eqnarray}
\noindent being $r_s=r_s(t)$ the event horizon radius. The radius $r_s$ represents a radial measure, nevertheless being a coordinate singularity. Moreover, if $r_s$ and the time $t$ are constants, the spacelike metric is given by
\begin{eqnarray}
    dl^2=r_s^2\bigl(d\theta^2 + \sin^2\theta d\phi^2\bigr).
\end{eqnarray}

The event horizon surface defined by $r_s$ allows us to state that any event occurring for $r>r_s$ is external to the BH. For the BH we are analyzing, the event horizon radius is given by the well-known result
\begin{eqnarray}\label{eq:shwarz}
    r_s=\frac{2GM}{c^2},
\end{eqnarray}
\noindent which allows us to write
\begin{eqnarray}\label{eq:entprodx}
    S_{bh}=\frac{4\pi}{T}\sigma r_s^2.
\end{eqnarray}

Observe that $\sigma r_s^2$ offers a measure of the energy/work associated with the event horizon radius. %Of course, the temperature $T$ also has \red{energy} \sout{kinetic} origin. Thus, the entropy $S_{bh}$ can be thought of as entirely due to kinetic energies.

The BH can capture mass/energy from the surroundings since it is a non-isolated system. Then, one assumes $M=M(t)$ leading the entropy \eqref{eq:entprodx} to be written as 
\begin{eqnarray}\label{eq:bh11}
    S_{bh}=\frac{16\pi G^2}{T c^4}\sigma M^2.
\end{eqnarray}

Considering the result \eqref{eq:bh11}, one can compute its entropy time derivative ($df/dt=\dot{f}$)
\begin{eqnarray}\label{eq:entprod1}
    \dot{S}_{bh}=\frac{32 \pi G^2}{T c^4}\sigma M\dot{M}.
\end{eqnarray}

If one assumes that $M$ is a slowly increasing function of $t$, then $dM/dt>0$ implies the entropy variation over $t$ grows according to $M$. Considering the Gouy-Stodola theorem, the equation \eqref{eq:entprod1} can represent the entropy produced only if it can be deduced from the reversible and irreversible work present in the system under study. Indeed, this is the case we show in the next section.

%%%%%%%%%%%%%%%%%
\section{Black Hole as a Gravitational Bubble}\label{sec:section_bubble}

Let us introduce a naive analogy between a BH and a soap bubble. The work to inflate a soap bubble depends on the surface tension $\sigma_{sb}$, defined as the force by unit length exerted by a stretched liquid covering layer (the membrane). The work is then stored as energy in the soap bubble membrane. 

The spherical shape of the soap bubble is obtained by minimizing the surface area for a given volume \cite{C.V.Boys.book}. Explicitly, the surface tension of a soap bubble can be defined as
\begin{eqnarray}\label{eq:sb}
    2\sigma_{sb}=(p_{int}-p_{ext})r_0,
\end{eqnarray}
\noindent where $p_{int}$ and $p_{ext}$ are the internal and external pressures, factor 2 considers the two sides of the membrane, and $r_0$ is the initial radius of the soap bubble. Observe that $p_{int}<p_{ext}$, implying the parameter $\sigma_{sb}$ is negative. 

Considering now the BH, when mass or energy falls inside the event horizon, it enhances $A_{eh}$ but the internal temperature of the BH, on the other hand, does not alter (significantly). Indeed, the internal temperature is near the vacuum temperature since it is inversely proportional to its mass. Then, the BH event horizon surface for the non-rotating and rotating cases, works similarly to the soap bubble membrane, storing the energy (work). Another similarity is that, as the soap bubble, the pressure inside of the BH decreases as its mass increases \cite{X.Calmet.F.Kuipers.Phys.Rev.D.104.066012.2021,Y.H.Wei.Phys.Lett.B672.98.2009,R.Campos.Delgado.Eur.Phys.J.C.82.272.2022,K.P.Chandra.JHEP.9.2.515.2023}. The similarities established above may provide interesting starting points for studying BHs as gravitational bubbles with surface tension. 

Recently, the surface tension for a BH was calculated \cite{L.Shu.K.Cui.X.Liu.Z.Liu.W.Liu.Fortschritte.der.Physik.67.3.1800076.2019}, corroborating the approach presented here. However, charged bubbles have a completely different behavior from the non-charged ones. The interplay between the surface tension and the electric field is crucial for understanding the bubble stability.  As the potential grows, the bubble becomes unstable at a critical voltage, leading to a Taylor cone and rupturing \cite{J.E.Hilton.A.van.der.Net.EPL86.2.24003.2009}. At this critical voltage, dynamic instabilities appear, and small bubbles are ejected from the original bubble \cite{J.E.Hilton.A.van.der.Net.EPL86.2.24003.2009}. Transposing this mechanism for a BH, the ejection of mass observed in some situations \cite{Cendes_2022} may have an origin in instabilities caused by the interplay between the gravitational tension and the electric field. At some critical value of the electric field, the BH ejects charged mass, diminishing the value of its electric field, and ceasing the charged mass ejection.

%%%%%%%%%%%%%%%%%5
\subsection{Non-Rotating Black Hole }

Considering the preceding discussion, one describes the gravitational surface tension associated with the event horizon surface as \eqref{eq:sb}, allowing us to write for the BH
\begin{eqnarray}\label{eq:pressures}
    2\sigma=(p_{int}-p_{ext})r_s.
\end{eqnarray}

In the present case, one assumes the only relevant external force acting is the gravitational field of the BH, which is a conservative force. On the other hand, the dominant internal force acting on the matter inside the black hole remains unknown. However, the existence of Hawking radiation, which leads to the BH evaporation, may suggest that this force has a non-conservative origin. Taking into account these considerations, note that one can write \eqref{eq:pressures} as
\begin{eqnarray}\label{eq:pressures2}
    2\sigma=(p_{int}-p_{ext})r_s=\left(\frac{F_{int}}{A_{eh}}-\frac{F_{ext}}{A_{eh}}\right)r_s=%\frac{1}{A_{eh}}\bigl(Fr_s-F_{out}r_s\bigr)=
    \frac{1}{A_{eh}}\bigl(W_{int}-W_{ext}\bigr),
\end{eqnarray}
\noindent where $W_{int}$ is the work due to non-conservative processes occurring inside the BH, whereas $W_{ext}$ accounts for the external conservative force. Then, identifying $W_{int}=W_i$ and $W_{ext}=W_r$ in the Gouy-Stodola theorem, one has 
\begin{eqnarray}\label{eq:pressures3}
   W_r-W_i=- 2\sigma A_{eh}=\sigma_{gb} A_{eh},
\end{eqnarray}
\noindent where $\sigma_{gb}>0$ since $\sigma<0$. The gravitational surface tension $\sigma_{gb}$ also emerges in different contexts, in which the effects of the gravitational potential emerge at large distances in the cosmic evolution \cite{C.Ortiz.Int.J.Mod.Phys.D29.16.2050115.2020,C.Ortiz.R.S.Khatiwada.Sci.Rep.13.10364.2023}. As aforementioned, gravitational surface tension is the energy contained in the event horizon, and, in the present case, this energy corresponds to the huge gravitational energy yielded by the gravitational potential. Then, the gravitational surface tension can be viewed as a manifestation of the curved spacetime \cite{C.Ortiz.Int.J.Mod.Phys.D29.16.2050115.2020,C.Ortiz.R.S.Khatiwada.Sci.Rep.13.10364.2023}. Applying the Gouy-Stodola theorem, one obtains 

\begin{eqnarray}\label{eq:entprod12}
    \dot{S}_{gb}=\frac{4\pi}{T}\bigl(2\sigma_{gb}r_s\dot{r}_s+\dot{\sigma}_{gb}r_s^2\bigr).%=\frac{32\pi G^2}{Tc^4}\sigma_{gb}M\dot{M},
\end{eqnarray}

The result \eqref{eq:entprod12} corresponds to the entropy yielded by a non-rotating BH under the gravitational bubble assumption. Comparing this development with \eqref{eq:entprod1}, one observes that Hawking-Bekenstein result is reached if $d\sigma_{gb}/dt\approx 0$, i.e., the gravitational surface tension varies slowly considering the BH time scale. In this case, one has
\begin{eqnarray}\label{eq:entprod2}
    \dot{S}_{gb}=\frac{8\pi}{T}\sigma_{gb}r_s\dot{r}_s=\frac{32\pi G^2}{Tc^4}\sigma_{gb}M\dot{M},
\end{eqnarray}
\noindent representing the same result as \eqref{eq:entprod1}, disregard some constant. However, in the present approach, the above result has a thermodynamic origin and represents the entropy production in the event horizon surface of the BH. 

Therefore, in the present model, the thermodynamic origin for the BH entropy is due to the gravitational surface tension, resulting in the entropy being naturally proportional to the event horizon surface. 

%%%%%%%%%%%%%%%%%5
\subsection{Rotating Black Hole }

Disregarding charge and spin effects, one considers a BH with angular momentum $J$ in a fixed background spacetime. This Kerr BH induces the Lense-Thirring effect - the precession of spacetime - leading the particles and light to rotate through the deformed spacetime. Therefore, the Kerr BH gravitational field depends also on its rotation. 

For the rotating BH, one writes \cite{J.M.Bardeen.B.Carter.S.W.Hawking.Commun.Math.Phys.31.161.1973,S.W.Hawking.Commun.Math.Phys.43.199.1975} (hereafter, for simplicity, $c=\hbar=G=1$)
\begin{eqnarray}\label{eq:tds1}
    dM=\frac{\kappa}{8\pi}dA + \Omega dJ,
\end{eqnarray}
\noindent where $\Omega=d\theta/dt$ is the angular velocity. The region between the outer event horizon and the outer ergosphere defines the ergosphere of such a BH. In the non-rotating case, of course, there is no ergosphere. The Boyer-Lindquist metric is written as \cite{R.H.Boyer.R.W.Lindquist.J.Math.Phys.8.265.1967}
\begin{eqnarray}
\nonumber ds^2 &=&-dt^2+\left(\frac{r^2+a^2cos\theta^2}{r^2+a^2} \right)dr^2+ (r^2+a^2\cos\theta^2 )d\theta^2 +\\
    &+& (r^2+a^2 )\sin^2\theta d\phi^2+\frac{2M}{r}\left[ \frac{(dt-a\sin^2\theta d\phi)^2}{1+a^2\cos^2\theta/r^2} + \frac{(1+a^2\cos^2\theta/r^2)dr^2}{1-2M/r+a^2/r^2}\right],
\end{eqnarray}
\noindent indicating the outer event horizon is given by
\begin{eqnarray}
    r_+=M+\sqrt{M^-a^2},
\end{eqnarray}
\noindent while the outer ergosphere is
\begin{eqnarray}\label{eq:oergon}
    r_+^{e}=M+\sqrt{M^2-a^2\cos^2\theta},
\end{eqnarray}
\noindent where $a=J/M$ is the Kerr parameter. When $a\rightarrow 0$, one recovers the Schwarzschild result, $r_+=r_+ ^e\rightarrow 2M=r_s$. Then, following $r_s$ also $r_+$ is a coordinate singularity in the Boyer-Lindquist line element. Furthermore, one also has the complimentary coordinate singularities  
\begin{eqnarray}
    r_-=M-\sqrt{M^-a^2},
\end{eqnarray}
\noindent while the inner ergosphere has the radius
\begin{eqnarray}
    r_-^{e}=M-\sqrt{M^2-a^2\cos^2\theta}.
\end{eqnarray}

Facing these various singularities, it is important to introduce the following question: whose coordinate singularity presented above represents the event horizon surface of the rotating BH? Since it is theoretically possible to extract work from the outer ergosphere using the Penrose process \cite{R.Penrose.R.M.Floyd}, one assumes the inner ergosphere represents the event horizon in the Schwarzschild sense of the non-rotating BH. Then, the gravitational surface is determined by the outer ergosphere.

Similarly to the non-rotating BH, the work for a massive particle outside the outer ergosphere, is due essentially to the gravitational field, being reversible. When the particle crosses this coordinate barrier given by \eqref{eq:oergon}, one considers once more the work is due to irreversible processes. Thus, for $r<r_+$, the work is due to irreversible processes.

For the non-rotating BH, the Gouy-Stodola theorem ensures $dS\propto dA$, where $T\equiv\kappa/8\pi$ is the proportionality constant. For the rotating BH, one writes
\begin{eqnarray}\label{eq:bh1}
    \Delta p= p_{int}-p_{ext}=2\pi\sigma,
\end{eqnarray}
\noindent where $\sigma$ is the surface tension upon the ergosphere. For small pressure variations, it is possible to write $\Delta p\approx p =p_{int}-p_{ext}<0$. In this case, the physical conditions are the same concerning the first law of thermodynamics \cite{J.M.Bardeen.B.Carter.S.W.Hawking.Commun.Math.Phys.31.161.1973}
\begin{eqnarray}
    TdS=dE+pdV,
\end{eqnarray}
\noindent where $E$ is the energy and $V$ the volume enclosed by the system. Analogously to equation \eqref{eq:pressures3}, one has
\begin{eqnarray}\label{eq:bh4}
    W_r-W_i-=-2\pi\sigma A_{eh},
\end{eqnarray}
\noindent where the right-hand side is proportional to the entropy. 

Comparing the first law and the Gouy-Stodola theorem, one obtains
\begin{eqnarray}\label{eq:firstlaw}
    dE+pdV=dW_r-dW_i.
\end{eqnarray}

For a massive particle subject to relativistic conditions, the energy infinitesimal variation is 
\begin{eqnarray}\label{eq:energmass}
    dE \propto dM,
\end{eqnarray}
\noindent for $r>r_+$. Consider the BH as an isothermal system where $dT/dt=0$ (or at least $dT/dt\approx 0$). The system is not isolated since matter and energy can enter the BH. Thus, there is a surface tension on the rotating gravitational bubble, and the time derivative of the angular momentum is not null
\begin{eqnarray}
    \frac{dJ}{dt}=\tau,
\end{eqnarray}
\noindent where $\tau$ is the torque. For an angular displacement $\theta$, the work done by the torque is given by
\begin{eqnarray}\label{eq:rf1}
   pdV= -\tau\theta = -\frac{dJ}{dt}\theta,
\end{eqnarray}
\noindent since $p$ is negative. The time derivative of the equation \eqref{eq:rf1} is given by
\begin{eqnarray}
    p\frac{dV}{dt}= -\frac{d^2J}{dt^2}\theta-\frac{dJ}{dt}\Omega.
\end{eqnarray}

The first term on the right-hand side is proportional to $d^2M/dt^2$ and $d^2r/dt^2$. Both terms can be neglected for small mass and radius variations compared to the BH total mass and event horizon radius. Therefore, one obtains
\begin{eqnarray}
    p\frac{dV}{dt}\approx -\frac{dJ}{dt}\Omega.
\end{eqnarray}

Using the result \eqref{eq:firstlaw}, one finally has
\begin{eqnarray}\label{eq:finalx}
    TdS\approx dM-\Omega dJ,
\end{eqnarray} 
\noindent corresponding to the well-known result from rotating black holes \cite{J.M.Bardeen.B.Carter.S.W.Hawking.Commun.Math.Phys.31.161.1973}. It is important to stress that taking into account the mass and event horizon radius variation, the term $d^2J/dt^2\theta$ means a thermodynamic correction for the result \eqref{eq:finalx}.

%Using $J=I\Omega=Mr_s^2\Omega$, one has
%\begin{eqnarray}
%    \frac{dJ}{dt}=r_s\bigl(\dot{M}r_s+2p_{||}\bigr)\Omega+I\dot{\Omega},
%\end{eqnarray}
%\noindent where $p_{||}=M(dr_s/dt)$. Then, the torque is given by
%\begin{eqnarray}\tau=r_s\bigl(\dot{M}r_s+2p_{||}\bigr)\Omega+I\dot{\Omega}.
%\end{eqnarray}
%Considering a constant angular velocity and small time variations for $r$, one has $\dot{\Omega}=0$ and  $dr_s/dt\approx 0$, implying
%\begin{eqnarray}
%\tau\approx\dot{M}r_s^2\Omega.
%\end{eqnarray}
%\begin{eqnarray}
%    pdV\approx -\tau\theta.
%\end{eqnarray}
%Then, disregarding $d^2M/dt^2$, one obtains
%\begin{eqnarray}\label{eq:trabtorque}
%    p\frac{dV}{dt}\approx -\frac{d(\tau\theta)}{dt}\approx -\frac{dM}r^2\Omega^2=-\frac{dJ}{dt}\Omega.
%\end{eqnarray}
%The results \eqref{eq:energmass} and \eqref{eq:trabtorque} can be replaced into Gouy-Stodola theorem to write 

%%%%%%%%%
\section{Merging Two Black Holes}\label{sec:merging}

%The entropy $S$ due to a BH with a constant radius $r_1+r_2$ is written as
%\begin{eqnarray}\label{eq:tds1}
%    TdS= \pi(r_1+r_2)^2d\sigma = \pi r_1^2 d\sigma + \pi r_2^2d\sigma + 2\pi r_1r_2d\sigma
%\end{eqnarray}

%In the simplest case, one assumes the gravitational surface tension is the same for both BH. Then, the entropy of each BH is written as
%\begin{eqnarray}\label{eq:tds2}
%    TdS_1+TdS_2=\pi r_1^2d\sigma+\pi r_2^2d\sigma.
%\end{eqnarray}

%Comparing the results \eqref{eq:tds1} and \eqref{eq:tds2}, one obtains for $d\sigma>0$
%\begin{eqnarray}\label{eq:tds3}
%    TdS_1+TdS_2<TdS.
%\end{eqnarray}

%If the surface tension of each BH is $\sigma_1$ and $\sigma_2$, then the sum of  entropies of each BH is written as
%\begin{eqnarray}\label{eq:tds4}
%    TdS_1+TdS_2=\pi r_1^2d\sigma_1+\pi r_2^2d\sigma_2
%\end{eqnarray}

%Assuming the surface tension of the BH with radius $r_1+r_2$ is a linear combination of the surface tension due to two different black holes, then it can be written as $\sigma=a_1\sigma_1+a_2\sigma_2$, $a_1,a_2\in \mathbb{R}$. Hence,
%\begin{eqnarray}\label{eq:tds5}
%TdS-(TdS_1+TdS_2)&=&\pi r_1^2(d\sigma-d\sigma_1) +\pi r_2^2(d\sigma-d\sigma_2) +2\pi r_1r_2 d\sigma.
%\end{eqnarray}

%For the cases where $d\sigma\geq d\sigma_{1,2}$, one has
%\begin{eqnarray}\label{eq:tds6}
%    TdS>TdS_1+TdS_2
%\end{eqnarray}

%For the case where $d\sigma<d\sigma_{1,2}$...

In the context of gas bubbles, non-spherical shapes can exist when their surfaces are covered with a close-packed monolayer of particles \cite{A.B.Subramanian.M.Abkarian.L.Mahadevan.H.A.Stone.Nature.438.930.2005}. In this case, the pressure variation depends on the local principal radii of curvature, $r_1$ and $r_2$, and the corresponding surface stress, $\sigma_1$ and $\sigma_2$ \cite{A.B.Subramanian.M.Abkarian.L.Mahadevan.H.A.Stone.Nature.438.930.2005}
\begin{eqnarray}\label{eq:dbubble}
    \Delta p=p_1-p_2=\frac{\sigma_1}{r_1}+\frac{\sigma_2}{r_2}<0.
\end{eqnarray}
\noindent where $p_1=p_{1int}-p_{1ext}=p_{1i}-p_{1e}$ and $p_2=p_{2int}-p_{2ext}=p_{2i}-p_{2e}$. Then, one writes 
%Let us assume the pressure in the interface (external) of two disjoint BH is the same: $p_{1ext}=p_{2ext}$. Thus, $\Delta p = p_{1int}-p_{2int}=p_{1i}-p_{2i}$ implies that when the event horizons interact with each other only internal processes are relevant to explain the merging of the two BH. Then, one writes \new{dúvida: a pressão external foi supostamente dita ser de origem do campo gravitacional do BH. Nesse sentido, não deveria colocar que há dois BH de memsma massa e área?}
\begin{eqnarray}\label{eq:dbubble}
    (p_{1i}-p_{1e})-(p_{2i}-p_{2e})= \frac{(W_{1i}-W_{1r})}{\pi r_1^3} - \frac{(W_{2i}-W_{2r})}{\pi r_2^3} = \frac{\sigma_1}{r_1}+\frac{\sigma_2}{r_2},
\end{eqnarray}
\noindent where $W_{1i}(W_{2i})$ stands for the work due to irreversible processes, and $W_{1r}(W_{2r})$ for the reversible ones in the BH 1(2). The last expression can be reorganized to furnishes \begin{eqnarray}\label{eq:dbubble}
    \frac{r_1^3}{r_2^3}= \frac{(W_{2i}-W_{2r})-\pi r_2^2\sigma_2}{(W_{1i}-W_{1r})+\pi r_1^2\sigma_1}.
\end{eqnarray}

Without loss of generality, one can assume $r_1/r_2\leq 1$, resulting in
\begin{eqnarray}
    (W_{2i}-W_{2r}) -(W_{1i}-W_{1r})\leq \pi r_1^2\sigma_1 + \pi r_2^2\sigma_2,  
\end{eqnarray}
\noindent or
\begin{eqnarray}
    (W_{1r}-W_{1i}) - (W_{2r}-W_{2i})\geq -(\pi r_1^2\sigma_1 + \pi r_2^2\sigma_2)\geq 0,  
\end{eqnarray}

Suppose both BH has the same internal temperature, $T_1=T_2=T$, near the vacuum temperature since the internal temperature of a BH is inversely proportional to its mass: the more massive the BH, the lower its temperature. From the Gouy-Stodola theorem, one has for an infinitesimal element
\begin{eqnarray}\label{eq:last1}
    T(dS_1-dS_2)=d(W_{1r}-W_{1i})-d(W_{2r}-W_{2i})\geq -(\pi r_1^2d\sigma_1 +  \pi r_2^2d\sigma_2)\geq 0
\end{eqnarray}
%Then, 
%\begin{eqnarray}\label{eq:last1}
%    T(dS_1-dS_2) \leq \pi r_1^2d\sigma_1 + % \pi r_2^2d\sigma_2  =T(dS_1+dS_2).
%\end{eqnarray}

On the other hand, for a BH with radius $R=r_1+r_2$, internal temperature $T$, and surface tension $\sigma=\sigma_1+\sigma_2$, one has
\begin{eqnarray}
    \Delta p= p_{int}-p_{ext}= \frac{\sigma}{R}= \frac{\sigma_1+\sigma_2}{r_1+r_2}.
\end{eqnarray}

Considering for this BH the event horizon area $A=\pi(r_1+r_2)^2$, one obtains
\begin{eqnarray}
    %F_{int}r_1-F_{ext}r_1+F_{int}r_2-F_{ext}r_2= 
    W_{int}-W_{ext}=W_i-W_r=\pi r_1^2\sigma+\pi r_2^2\sigma+2\pi r_1r_2\sigma.
\end{eqnarray}

For infinitesimal variations, the Gouy-Stodola theorem
\begin{eqnarray}
    TdS=dW_r-dW_i,
\end{eqnarray}
\noindent combined with the result \eqref{eq:last1}, can be used to write the following inequality
\begin{eqnarray}
    TdS -T\bigl(dS_1+dS_2\bigr)=-(\pi r_2^2d\sigma_1+\pi r_1^2d\sigma_2+2\pi r_1r_2d\sigma)>0,
\end{eqnarray}
\noindent since $d\sigma_1,d\sigma_2<0$. Therefore, one obtains from the present approach the well-known result
\begin{eqnarray}\label{eq:hawking10}
    TdS>TdS_1+TdS_2
\end{eqnarray}
ensuring the entropy always increases. The inequality \eqref{eq:hawking10} corroborates the original development due to Hawking \cite{S.W.Hawking.Commun.Math.Phys.43.199.1975}. However, notice the inequality is obtained from thermodynamic considerations assuming the validity of the Gouy-Stodola theorem and the gravitational bubble model. 

%%%%%%%%%
\section{Discussion}\label{sec:disc}

Thermodynamic laws live in the core of physics and, in principle, these laws could be applied to every physical system, despite the complexity or number of variables. It is important to highlight that using analogies (such as gravitational bubbles presented here) can be a double-edged sword. While it makes it easier to understand, it may also be considered an oversimplification, especially in a field where mathematical and physical precision are crucial.

The Gouy-Stodola theorem, originally derived for mechanical systems under simple assumptions, is applied in this context, assuming the black hole can be described as a gravitational bubble. As in the soap bubble, where energy information is restricted to the membrane, in the gravitational bubble model for the BH, the gravitational surface tension retains all the information about the BH energy. Then, the only relevant quantity for the present model is the event horizon surface and its possible relations with other relevant physical quantities (such as mass, angular momentum, and charge).

This theorem provides a thermodynamic basis for the entropy-area relationship in black holes. Using the gravitational surface tension to model the BH, the Gouy-Stodola theorem coherently explains the entropy production in non-rotating and rotating black holes, and it can also be extended to scenarios involving black hole mergers. 

For the non-rotating BH, the relationship between the event horizon area, and the entropy is obtained by employing this theorem, corroborating the well-known result of Hawking and Bekenstein: the black hole entropy is proportional to the event horizon surface. The entropy-area relation is also derived for the non-charged, rotating BH. 

Finally, merging two black holes implies the resulting entropy (or the event horizon area) is greater than the sum of the original entropy of each black hole. 

The case of a charged rotating BH is presently under study since the equations governing the dynamics of a charged bubble are more intricate than the non-charged one. 

%%%%%%%%%%%%%%to %%%%
\section*{Acknowledgments}

SDC and RHL thank UFSCar for the financial support.

%%%%%%%%%%%%%%%%%%

\end{document}